\documentclass[12pt]{article}
\usepackage{setspace,graphicx,srcltx,url,lscape,textcomp,longtable,amssymb,amsthm,enumerate,enumitem,amsmath,color,booktabs,array,calc,epigraph,microtype,lipsum,moresize,todonotes,xcolor,subcaption,float,array,titlesec,bbm}
\usepackage{natbib}
\usepackage{multirow}
\usepackage{url} 

\newcommand{\blind}{1}

\begin{document}

\def\spacingset#1{\renewcommand{\baselinestretch}%
{#1}\small\normalsize} \spacingset{1}


\if1\blind
{
  
  \title{\bf Bang the Can Slowly: An Investigation into the 2017 Houston Astros }
  
  
  \author{Ryan Elmore \\
  Department of Business Information and Analytics \\
  Daniels College of Business, University of Denver \\
	and \\
	Gregory J. Matthews \\Department of Mathematics and Statistics\\ Loyola University Chicago
    }
  \maketitle
  \newpage
} 

\if0\blind
{
  \newpage
  \bigskip
  \bigskip
  \bigskip
  \begin{center}
    {\LARGE\bf Stolen Signs}
\end{center}
  \medskip
} \fi

\bigskip
\begin{abstract}
   This manuscript is a statistical investigation into the 2017 Major League Baseball scandal involving the Houston Astros, the World Series championship winner that same year. The Astros were alleged to have stolen their opponents' pitching signs in order to provide their batters with a potentially unfair advantage. This work finds compelling evidence that the Astros on-field performance was significantly affected by their sign-stealing ploy and quantifies the effects. The three main findings in the manuscript are: 1) the Astros' odds of swinging at a pitch were reduced by approximately 27\% (OR: 0.725, 95\% CI: (0.618, 0.850)) when the sign was stolen, 2) when an Astros player swung, the odds of making contact with the ball increased roughly 80\% (OR: 1.805, 95\% CI: (1.342, 2.675)) on non-fastball pitches, and 3) when the Astros made contact with a ball on a pitch in which the sign was known, the ball's exit velocity (launch speed) increased on average by 2.386 (95\% CI:  (0.334, 4.451)) miles per hour.  
\end{abstract}


\noindent%
\vfill

\newpage
\spacingset{1.45} 
\section{Introduction}
\label{sec:intro}
Prior to each pitch in a baseball game, the pitcher and the catcher will communicate information on the type of pitch (e.g., fastball, curveball, etc.) to be thrown. The brief exchange ensures that the catcher knows exactly what to expect from the pitcher in order to react appropriately. This is important because different types of pitches may look very similar leaving the pitchers hand, however, they may exhibit wildly different behavior as they travel toward home plate.  

The standard method of communication relies on the catcher flashing hand signals between his legs as he is squatting behind home plate. In this manner, the pitcher can see what is being relayed, but the sign is usually hidden from the batter's view. The pitcher will confirm or change a sign of a particular pitch by nodding his head ``yes'' or ``no'', respectively, to the catcher. Once they are in agreement, the pitcher winds up and throws the pitch. 

It is important that the catcher obscure the signs from the batter so that the batter does not know the type of pitch that is coming next.  A pitcher relies on this uncertainty in order to confuse the batter and ultimately achieve a more favorable outcome for his team.  On the other hand, if the batter knows the type of pitch that is coming, he may reduce the pitcher's considerable advantage.  One way for the batter to learn the pitching sign is for the batter's team to steal the signs from the catcher, decode the signals, and then somehow share the decoded signal with the batter. 
During the 2017 Major League Baseball (MLB) season, the Houston Astros are alleged to have implemented a elaborate sign-stealing scheme. In 2019, Mike Fiers, a pitcher for the Houston Astros during their 2017 World Series championship run, claimed that his former team was stealing signs by using a camera in center field, \cite{nytimes_fiers}. The information in these stolen signs was relayed to players by banging a baseball bat against a trash can \citep{athletic_rosenthal_drellich}, referred to here as a ``bang''. In this particular scheme, a bang indicated to the batter that the upcoming pitch would be an off-speed pitch such as a curveball or a slider. The absence of a bang is inconclusive; it could indicate a fastball or that they simply could not decode the sign.

Ultimately, MLB punished the Houston Astros by suspending their manager, A.J. Hinch, and general manager, Jeff Lunhow, for one year.  Additionally, the Astros were fined \$5 million and their first and second round draft picks were taken away for the 2020 and 2021 amateur drafts. This was a substantial penalty, and meant to dissuade future teams from impacting their games in a similar manner.\footnote{The Boston Red Sox were also swept up in the scandal by way of Alex Cora, formerly the bench coach of the Houston Astros, who coached the Red Sox in 2018 to a World Series Championship.  He is alleged to have developed a sign stealing system in Boston, and was subsequently fired by the Red Sox.}

However, not everyone agrees on the effects of stealing pitching signs during an MLB game. In one particularly bizarre exchange during a press conference on February 13, 2020, the owner of the Astros, Jim Crane, was quoted as saying ``\textbf{Our opinion is this didn't impact the game}. We had a good team. We won the World Series and we'll leave it at that.'' In that same press conference, less than a minute later, he is also quoted as saying, ``I didn't say it didn't impact the game.'' See \cite{Axisa2019} for more information on this press conference.

Others, have found many striking examples of different aspects of the game that appear to show that that Houston benefited from sign stealing.  For instance, \cite{sawchik538} notes a large decrease in strikeout rates for the Astros from 2016 to 2017.  \cite{stark2020} reaches similar conclusions to \cite{sawchik538}, while also noting large differences in strike out rates between home and away games, notable drops in swing rates for some players from 2016 to 2017, and large increases in slugging from 2016 to 2017.  In addition, they found several players had dramatic decreases in their strikeout rates, notably George Springer, Carlos Correa, Evan Gattis, and Jake Marisnick. \cite{ArthurBP} found intraseason improvements before and after he first observed the implementation of the Astros' scheme (May 19th) in both swinging strike rates and swinging at likely balls.  \cite{ArthurBP} also notes an improvement in exit velocity. Alternatively, \cite{LindberghRinger} found little evidence that the Astros gained much by stealing signs when looking at overall performance.  

While there has been quite a bit of analysis on the on-field effects of the Astros' sign stealing, it is still largely an open question as to whether or not there were on-field improvements because of the sign stealing. In addition, if stealing signs did lead to improvement, what types of improvements were observed and can the magnitude of these improvements be quantified. In this paper, we attempt to erase any ambiguity related to the efficacy of the Astros' sign-stealing scheme during the 2017 season. In other words, we address whether or not their scheme affected on-field performance during the 2017 season and quantify their impacts where appropriate.

 The paper is outlined as follows. We describe the three data sets that we used for this analysis in Section \ref{sec:data}. Next, we present our strategy and methodology in Section \ref{sec:methods}. Our results are presented in Section \ref{sec:results} and finally we close with our concluding remarks and a discussion in Section \ref{sec:conc}.

\section{Data}
\label{sec:data}

In the analysis that follows, we rely on, and leverage the strengths from, three distinct data sources: Statcast, Pitch Info, and Bangs. Each source is described in the following paragraphs.  

Statcast is Major League Baseball's ball and player tracking system that has been in every MLB park since the 2015 season. A new version of Statcast, Version 2.0, is set to be released in 2020. The Statcast V1.0 system has two data collection components: (1) Trackman Doppler Radar that tracks baseball events and Chryon Hego Cameras that track player movements. In the first three seasons alone, 2.1 MM pitches and 400K balls in play were tracked. Examples of variables that are available in Statcast include a hit's launch speed (exit velocity), pitch classification, pitch spin rate, among a host of additional measurements. See \citet{statcast} for additional information related to MLB's Statcast application programming interface (API). 

Although pitch classification is available in Statcast, we relied on Pitch Info \cite{PitchInfo} data for classifications rather than Statcast. Pitch Info is regarded as the most accurate classification system in terms of pitch group classification in the industry. In addition, we utilize Pitch Info's derived variable, referred to as called strike probability (CSP) as a covariate in our analysis. CSP is an estimate of the probability that a pitch will be called a strike. Full details related to the CSP model can be found here \cite{Judge2015}.

Finally, we merge the previous two data sets with the so-called Bangs data. These data contain information on whether or not a measurable, auditory signal was present prior to pitches on a selection of Astros' at-bats during a subset of their Astros' 2017 season's home games. The signal, if present, was the result of banging a hard object on a metal trash can. The data were compiled by Tony Adams, a self-described Astros fan, and are publicly available on his website \citep{Adams2019}. The data are essentially a combination of data obtained from Major League Baseball's Statcast API along with video of Houston Astros' games from Youtube. Adams matched timestamps from the MLB Statcast data to the game video, and produced a spectrogram to represent the audio before and after all pitches in his study. The spectrogram of the auditory footprint of each of these pitches was used to identify when bangs were present prior to a pitch.  

The original Bangs data set has $n = 8274$ observations, however, we had to remove a number of observations. For instance, we removed observations with missing pitch ids, bunts/bunt attempts, ambiguous pitch groups, among other oddities. The resulting data set has a total of $n = 8201$ observations and are available on our github site \cite{CodeGit}.

\begin{figure}[tbph]\centering
\includegraphics[width=0.95\textwidth]{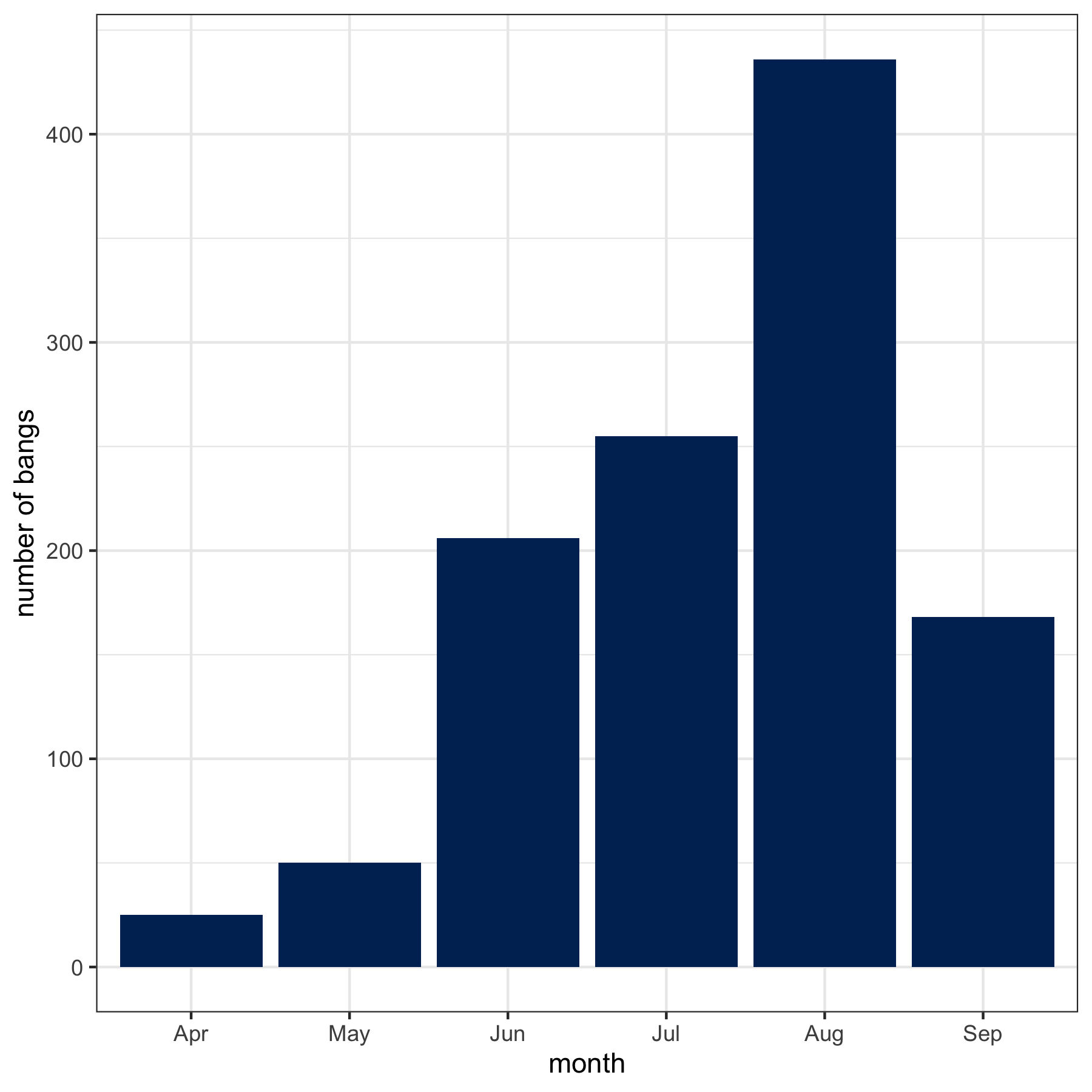}
\caption{This figure shows the proportion of bangs per month in our combined data set. It is clear that the Astros became more confident in their sign stealing system as the season progressed.}
\label{fig:bangs-by-month}
\end{figure}

After combining the three data sets described above, we summarized the bangs looking for trends over time, as well as trends among the individual players. Figure \ref{fig:bangs-by-month} \citep{teamcolors} shows the number of bangs per month relative to the total number of at bats in our data set. It is easy to see that the proportion of bangs per month increased steadily through the first five months of the 2017 season and then dropped off in September. This suggests that the Astros became more confident as the season progressed and/or it took them a while to refine their sign-stealing system.

In addition to the bangs per month, we looked at the number of pitches that included a bang for various players. Specifically, we selected nine players with the most pitches thrown to in our combined data set and looked at how many plate appearances included a bang and how many did not. Clearly, it would appear that certain players preferred to not hear the bang alert (e.g., José Altuve and Josh Reddick) at the same rate as the rest of the team.  

\begin{figure}[tbph]\centering
\includegraphics[width=0.95\textwidth]{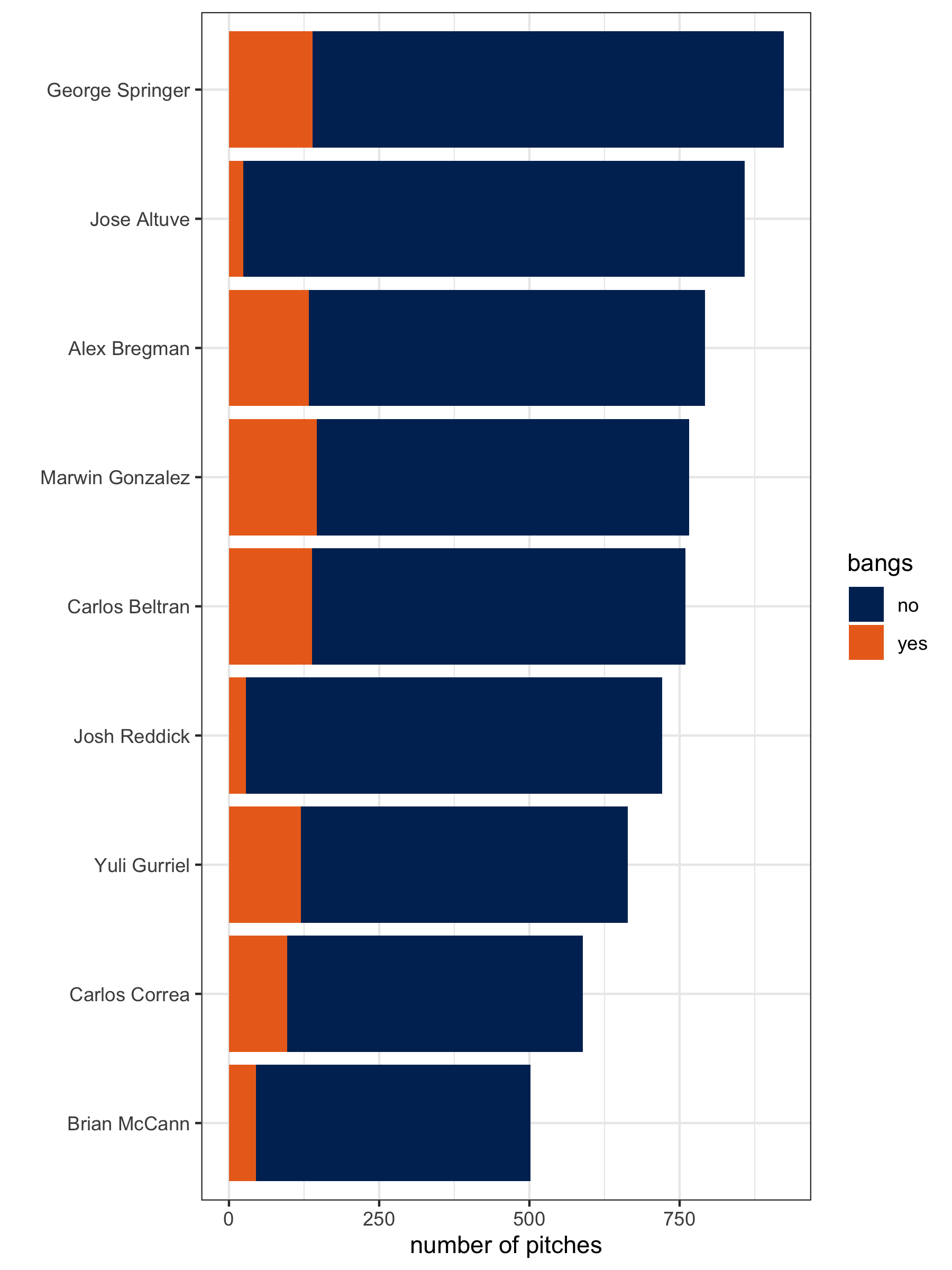}
\caption{This figure shows number of pitches and bangs for the nine players on the Astros who had the most at bats in our final data set.}
\label{fig:bangs-by-player}
\end{figure}


Next, we looked at the prevalence of bangs for different types of pitch groups, as defined in Pitch Info, see Table \ref{table_pitch_group_vs_bang}.  It is immediately clear that there is a relationship between pitch group and the incidence of a bang. In fact, a $\chi^2$-test of independence yields a p-value of $<2.2 \times 10^{-16}$, strongly rejecting the null hypothesis of independence between the two variables. In particular, offspeed pitches such as change-ups (CH), curveballs (CU), and sliders (SL) show a bang prior to the pitch at rates of 23.7\%, 27.6\%, and 26.8\%, respectively. On the other hand, fastballs (FA) only had a bang prior to the pitch on 2.3\% of the pitches. So while the Astros' method was not perfect, it is obvious that information such as an upcoming offspeed pitch was being transmitted to the batter via the trash can banging system.  

\begin{table}[ht]
\centering
\caption{The number of pitches of Pitch Info pitch group and the incidence of bangs. The percentages correspond to the bang prevalence conditioned on pitch type category.}
\label{table_pitch_group_vs_bang}
\renewcommand{\arraystretch}{1.25}
\begin{tabular}{llrrrr}
\toprule
 &  & \multicolumn{4}{c}{\bf{Pitch Type}} \\ \cline{3-6}
 & & \multicolumn{1}{c}{Change-up} & \multicolumn{1}{c}{Curveball} & \multicolumn{1}{c}{Fastball} & \multicolumn{1}{c}{Slider} \\ 
  \hline
  \multirow{2}{4em}{\bf{Bangs}} & No & 756 (76.3\%) & 707 (72.4\%) & 4128 (97.7\%) & 1470 (73.2\%) \\ 
   & Yes & 235 (23.7\%) & 270 (27.6\%) &  97 (2.3\%) & 538 (26.8\%) \\ 
   \bottomrule
\end{tabular}
\end{table}

Finally, Table \ref{table_swing_vs_bang} shows the relationship between swinging at a pitch and whether or not there was a bang preceding the pitch. Given the presence of a bang, an Astros' player swung 40.5\% of the time as opposed to  46.2\% when there was no bang. This translates to an odds ratio of 0.793 (95\% CI: 0.69678,  0.9023; p-value $3.706 \times 10^{-4}$) , which indicates an approximate 21\% reduction in the odds of swinging given a bang relative to when there was no banging before the pitch. While there is certainly a significant relationship between swinging and the presence of a bang, there are many additional variables that can confound this relationship. We will explore this in more detail in the following section. 

\begin{table}[ht]
\centering
\caption{The number of pitches with bangs by the number of swings.  Percentages correspond to the percentage of swings given bangs/no bangs on the pitch.}
\renewcommand{\arraystretch}{1.25}
\begin{tabular}{llrr}
  \toprule
 &  & \multicolumn{2}{c}{\bf{Swing}} \\ \cline{3-4} 
 & & \multicolumn{1}{c}{No} & \multicolumn{1}{c}{Yes} \\
  \hline
  \multirow{2}{4em}{\bf{Bangs}} & No & 3798 (53.7\%) & 3263 (46.2\%) \\
  & Yes & 678 (59.5\%) & 462 (40.5\%) \\ 
   \bottomrule
\end{tabular}
\label{table_swing_vs_bang}
\end{table}



\section{Methodology}
\label{sec:methods}
In this section, we describe our approach to analyzing several possible effects of stealing signs on measurable, in-game quantities. We employ a layered, or conditioned, approach to our analysis in that we start with an investigation into the effect on swinging at a given pitch, refine our data and methods to examine whether a swing results in making contact with the ball, and finally we investigate the quality of contact through exit velocity at our most granular level. 

In their most general form, we are simply fitting generalized linear mixed models (GLMM) in each case \citep{mcculloch2014generalized}. The particular GLMM is determined by its response variable and covariates, however, they can be defined in general using Equation \eqref{eq:glmm} by 
\begin{align}
    \label{eq:glmm}
\boldsymbol{\eta} = \mathbf{X}\boldsymbol{\beta} + \mathbf{Z}\mathbf{b}.
\end{align}
Here $\boldsymbol{\eta}$ is referred to as the linear predictor and is related to the response of interest through a link function $g(\cdot)$, $\mathbf{X}$ is a matrix of covariates, $\boldsymbol{\beta}$ is a vector of parameters (fixed-effects), $\mathbf{Z}$ is the random effect design matrix, and $\mathbf{b}$ is the vector of random effect parameters. The specific GLMMs that we utilize are defined by the response variables of interest, the link $g(\cdot)$, covariates, and random effect terms. The specific terms are given explicitly in the following subsections.


\subsection{Swing Model}
\label{ssec:swing_model}

We developed our first GLMM in order to estimate the effect of stealing a pitching sign on swinging at the subsequent pitch. Specifically, let $Y_{ij}$ be an indicator for the $j^{th}$ player swinging at the $i^{th}$ pitch, $i = 1, \cdots, n_j$ and $j = 1, \cdots, N_{players}$. The link function in this situation is the well known $\text{logit}(\pi_{ij}) = \text{log}\frac{\pi_{ij}}{1-\pi_{ij}}$, where $\pi_{ij} = P(Y_{ij} = 1)$ conditioned on an indicator variable for the presence or absence of a bang, while controlling for pitch type ({\em i.e.}, fastball or not), CSP, pitch count (as a factor). By pitch count, we mean the number of balls and strikes that a batter has while facing the current pitch. Additionally a random slope term for each batter is included in the model. 


\subsection{Contact Model}
\label{ssec:contact_model}

Given that a swing occurred, we now focus our attention on whether or not contact was made with the ball. In other words, did the batter actually hit the ball (not necessarily hit in play) that was pitched when he swung? In this case, we define another binary response variable to be $Y_{ij} = 1$ if the $j^{th}$ player made contact with the ball on the $i^{th}$ pitch, $i = 1, \cdots, n_j$ and $j = 1, \cdots, N_{players}$. Similar to the situation given in the previous subsection, the link function in this model is the $\text{logit}(\pi_{ij})$ where $\pi_{ij}$ is the probability of making contact given fixed-effect covariates defined by CSP, an indicator variable for a fastball, an indicator variable for the presence of a bang, and an interaction term between fastball and bang. We include random intercepts for {\em both} pitcher and batter as well as a random slope for bangs for battter.  


\subsection{Exit Velocity Model}
\label{ssec:ev_model}

\begin{figure}[tbph]
\centering
\includegraphics[width=0.95\textwidth]{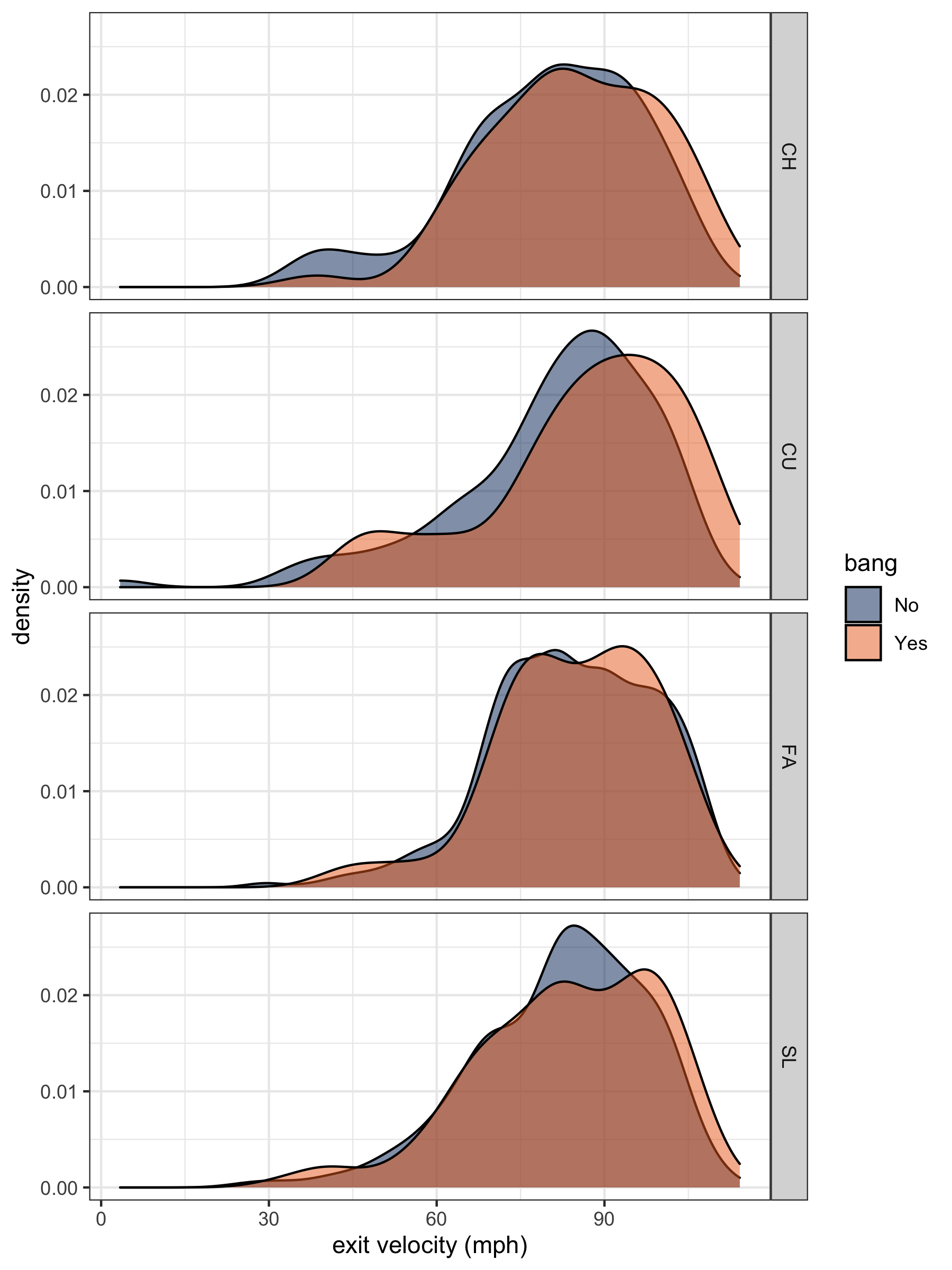}
\caption{This figure shows exit velocity by four different pitch types. The pitch types are change up (CH), curveball (CU), fastball (FA), and slider (SL). The non-fastball pitchtypes are all off-speed pitches.}
\label{fig:ev}
\end{figure}

Finally, the hierarchical nature of our modeling process leads us to examining a hit's exit velocity (EV) based on the presence of bangs prior to the pitch. Given that the batter swung at the pitch {\em and} made contact with the ball, was exit velocity impacted by stealing the pitching sign? In other words, are hits better given knowledge of an upcoming pitch? A preliminary look at EV by pitch type (see Figure \ref{fig:ev}) suggests that the average EV might be higher when a bang is present on off-speed pitches. 

The specific GLMM model in the case is a simpler form than the two described above. We treat $Y_{ij}$, the exit velocity of the ball leaving the bat, as a continuous variable. Therefore, the link function is simply the identity and, hence, we use a standard linear mixed-effects model. Covariates in this model include CSP, an indicator variable for fastball, an indicator variable for the presence of a bang, and random intercepts for both pitcher and batter.






\section{Results}
\label{sec:results}





\subsection{Swing Model}

The results of fitting the model described in Section 3.1 are give in Table \ref{table_swing_model_results}. Of primary interest, the coefficient estimate for the bangs indicator variable is -0.3219 (95\% CI: (-0.482, -0.163), p-value: $7.59 \times 10^{-5}$) indicating that when there were bangs prior to a pitch, the batter was significantly less likely to swing at that pitch relative to pitches with no bangs present. When all other covariates are help constant, the odds ratio for the probability of swinging comparing bangs to no bangs is 0.725 (95\% CI: (0.618, 0.850)). This translates to an approximate 27.5\% reduction in the odds of swinging in the presence of a bang. This is conclusive and statistically significant evidence that on-field behavior was directly affected by stealing the pitcher's sign. That is, the act of banging on a drum prior to a pitch (to indicate the ensuing pitch type) provided significant information to the batter causing him to swing less often.

\begin{table}[ht]
\centering
\caption{Fixed effect estimates for the swing model with the effect of bangs in bold. Note that the reference pitch count (PC) category is zero balls and zero strikes, denoted PC:0-0. With the exception of CSP (called-strike probability), all of the terms are indicator variables.}
\renewcommand{\arraystretch}{1.25}
\begin{tabular}{rrrrr}
  \toprule
 Term & Estimate & Std. Error & Z Statistic & p-value \\ 
  \hline
  Intercept & -2.47 & 0.10 & -25.53 & 0.00 \\ 
  CSP & 2.50 & 0.07 & 35.14 & 0.00 \\ 
  $I_{\{\textrm{Fastball}\}}$ & 0.06 & 0.06 & 1.06 & 0.29 \\
  $I_{\{PC:0-1}\}$ & 1.17 & 0.09 & 12.69 & 0.00 \\ 
  $I_{\{PC:0-2}\}$ & 1.90 & 0.13 & 15.20 & 0.00 \\ 
  $I_{\{PC:1-0}\}$ & 0.54 & 0.09 & 5.74 & 0.00 \\ 
  $I_{\{PC:1-1}\}$ & 1.47 & 0.10 & 15.21 & 0.00 \\ 
  $I_{\{PC:1-2}\}$ & 2.17 & 0.10 & 20.76 & 0.00 \\ 
  $I_{\{PC:2-0}\}$ & 0.53 & 0.14 & 3.81 & 0.00 \\ 
  $I_{\{PC:2-1}\}$ & 1.51 & 0.12 & 12.86 & 0.00 \\ 
  $I_{\{PC:2-2}\}$ & 2.49 & 0.11 & 22.61 & 0.00 \\ 
  $I_{\{PC:3-0}\}$ & -1.13 & 0.28 & -4.01 & 0.00 \\ 
  $I_{\{PC:3-1}\}$ & 1.34 & 0.17 & 8.02 & 0.00 \\ 
  $I_{\{PC:3-2}\}$ & 2.46 & 0.14 & 18.03 & 0.00 \\ 
  $I_{\{Bang\}}$ & {\bf -0.32} & 0.08 & -3.96 & 0.00 \\ 
   \hline
   $\sigma_b$ &  0.2079  & & &\\
   \bottomrule
\end{tabular}
\label{table_swing_model_results}
\end{table}

It is worth discussing why a player might swing more often given that he knows a fastball is being pitched. Simply put,  fastballs are easier to hit. \citet{fastballs} states that the MLB batting average is approximately 20\% - 40\% higher on fastballs and that the off-speed pitches lead to less contact. In other words, it is extremely useful for the batter to know that a harder-to-hit off-speed pitch is coming.



\subsection{Contact Model}
Next, we looked at estimating the probability of making contact with a pitch given that a player swung the bat. If a player has prior information about an upcoming pitch (e.g., the pitch is not a fastball), then we posit that the player is less likely to miss the ball entirely given the knowledge of pitch-type.

Prior to fitting this model, pitches where the player did not swing were removed from the data set, along with several unrelated and rare results, e.g. batter or catcher interference. This leaves 3725 observations. Given that a swing occurs, we defined ``contact'' on a pitch as any result other than a swing and miss occurs. Therefore, a ball put in play, regardless of whether or not they made an out, a foul ball, or a home run are all treated equally as ``contacts''.

\begin{table}[ht]
\centering
\caption{The fixed effect estimates for the contact model with the effect of banging on a metal can in bold.}
\renewcommand{\arraystretch}{1.25}
\begin{tabular}{rrrrr}
  \toprule
 Term & Estimate & Std. Error & Z Statistic & p-value \\ 
  \hline
  Intercept & -0.20 & 0.14 & -1.46 & 0.15 \\ 
  CSP & 1.90 & 0.12 & 16.37 & 0.00 \\ 
  $I_{\{\textrm{Fastball}\}}$ & 0.97 & 0.10 & 9.61 & 0.00 \\ 
  $I_{\{Bang\}}$ & {\bf 0.59} & 0.22 & 2.68 & 0.01 \\ 
  $I_{\{\textrm{Fastball}\}}$*$I_{\{Bang\}}$ & {\bf -1.19} & 0.45 & -2.64 & 0.01 \\ 
   \bottomrule
\end{tabular}
\label{table_contact_model}
\end{table}





A summary of the results of fitting the model described in Section 3.2 are given in Table \ref{table_contact_model}. Notice that both the indicator variable for banging on the can and its interaction effect with the fastball indicator are both significant. For this reason, we interpret fastballs and off-speed pitches separately.  First, the estimated effect size for bangs is 0.591 (95\% CI: 0.294, 0.984) on off-speed pitches.  This corresponds to an odds ratio of 1.805 (Bootstrapped 95\% CI: 1.342, 2.675, see \citet{Efron}). In other words, given that a player swings at the pitch, the odds of making contact in the presence of a bang (the pitching sign is known) are about {\it 80\% higher} than the odds of making contact when a bang is not present. Once again, we have found evidence that banging on a trash can prior to a pitch, or stealing the pitcher's sign, demonstrably affects on-field performance.  

Next, we will consider fastballs. The coefficient for bangs when the pitch is a fastball is -0.603. This corresponds to an odds ratio of 0.547 (Bootstrapped 95\% CI: 0.227, 1.774). While this odds ratio is not significantly different than one, an odds ratio less than one here would mean that a player is {\it less likely} to make contact in the presence of a bang. A bang prior to a pitch that turns out to be a fastball is actually a mistake on the part of the sign thief (i.e., the batter is expecting an off-speed pitch). It would not be surprising to see that batters have a harder time making contact on a pitch when they were expecting a pitch that was different than what was actually thrown.

\begin{figure}[tbph]\centering
\includegraphics[width=0.9\textwidth]{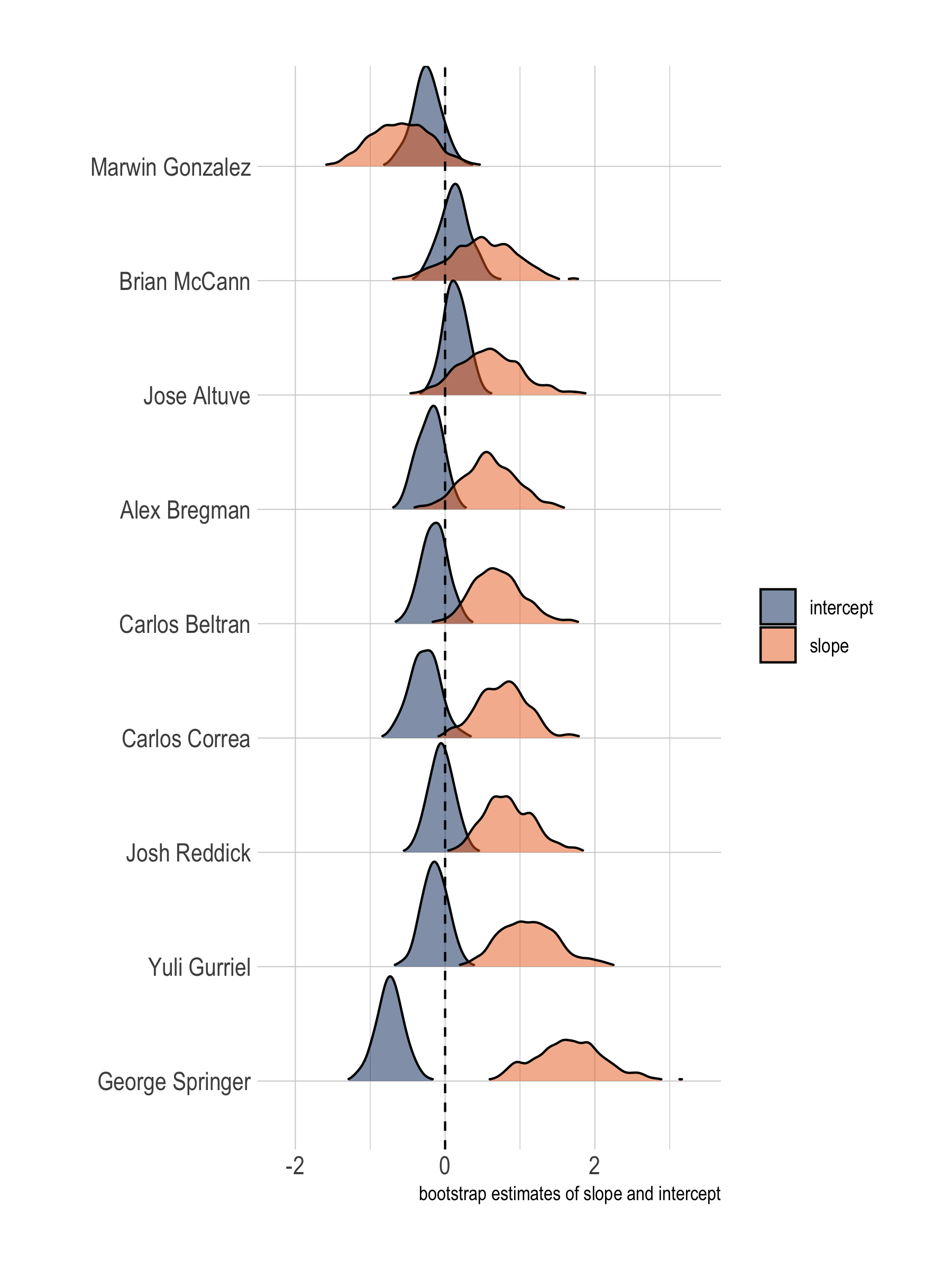}
\caption{This figure shows the player-specific bootstrap sampling distributions of the intercept and slope (associated with the indicator variable for banging) parameters for the nine players on the Astros who had the most at bats in our final data set.}
\label{fig:slope-intercept}
\end{figure}

Table \ref{table_re_contact} shows player-specific odds ratios for off-speed pitches, along with corresponding 95\% confidence intervals. The estimates are derived from the player-specific slopes associated with the indicator of bangs for all of the Astros' players having seen a pitch in our data set. Furthermore, the bootstrap sampling distributions of the estimates for the nine players who faced the most pitches are displayed in Figure \ref{fig:slope-intercept}. It is clear from Table \ref{table_re_contact} that any increase in the probability of contact on off-speed pitches given a bang prior to the pitch was highly variable across players.  In fact, of the twenty players included in our data set, ten of them do not exhibit a statistically significant increase in the odds of making contact on off-speed pitches given a swing (i.e., their confidence interval contains 1).

\begin{table}[ht]
\centering
\caption{Player-specific odds ratios for the effect of banging on a steel can prior to a pitch (i.e., stealing the pitch sign) on making contact along with 95\% bootstrap intervals for off-speed pitches.}
\renewcommand{\arraystretch}{1.25}
\begin{tabular}{lcc}
  \toprule
{\bf Name} & {\bf Odds Ratio} & {\bf Confidence Interval} \\ 
  \hline
George Springer & {\bf 3.810} & (2.042, 12.864) \\ 
  Yulieski Gurriel & {\bf 2.586} & (1.485, 7.279) \\ 
  Jonathan Davis & 2.416 & (0.869, 12.011) \\ 
  Jacob Marisnick & {\bf 2.377} & (1.250, 5.765) \\ 
  Evan Gattis & {\bf 2.050} & (1.017, 4.541) \\ 
  Josh Reddick & {\bf 2.010} & (1.368, 4.722) \\ 
  Max Stassi & {\bf 1.898} & (1.326, 3.376) \\ 
  Carlos Correa & {\bf 1.864} & (1.079, 4.182) \\ 
  Carlos Beltran & {\bf 1.848} & (1.074, 4.146) \\ 
  Juan Centeno & 1.794 & (0.777, 4.035) \\ 
  Alex Bregman & 1.774 & (0.891, 3.771) \\ 
  Derek Fisher & 1.751 & (0.742, 4.259) \\ 
  Norichika Aoki & {\bf 1.750} & (1.122, 3.52) \\ 
  Cameron Maybin & 1.737 & (0.763, 3.907) \\ 
  Anthony Kemp & {\bf 1.680} & (1.008, 2.747) \\ 
  Jose Altuve & 1.609 & (0.769, 4.547) \\ 
  Andrew Reed & 1.547 & (0.275, 2.998) \\ 
  Brian McCann & 1.480 & (0.646, 3.772) \\ 
  Tyler White & 1.093 & (0.231, 2.335) \\ 
  Marwin Gonzalez & 0.671 & (0.262, 1.199) \\ 
   \bottomrule
\end{tabular}
\label{table_re_contact}
\end{table}

The ten remaining players George Springer, Yulieski Gurriel, Jacob Marisnick, Evan Gattis, Josh Reddick,  Max Stassi, Carlos Correa, Carlos Beltran, Norichiak Aoki, and Anthony Kemp, on the other hand, all exhibited significant increases in their respective odds of contact given the presence of bangs prior to the pitch. Nine of these ten remaining players had increases in their odds of contact on off-speed pitches ranging from 68\% (Anthony Kemp) to 159\% (Yulieski Gurriel).  However, one player, George Springer, seems to have benefited much more than the other players with an estimated 281\% increase in the odds of contact on an off-speed pitch when a pitch was preceded by a bang.  

To put this in perspective, in our data set we have 390 records of George Springer swinging and 135 of those swings were at off-speed pitches. Ninety-five of these swings at off-speed pitches were not preceded by bangs and 31 of these swings resulted in no contact for a swing-and-miss rate of 32.63\% (95\% CI: 23.57\%, 43.12\%). Out of the remaining 40 swings at off-speed pitches that were preceded by bangs {\it only 2 resulted in no contact}.  This corresponds to a miss rate of only 5\% (0.87\%, 18.21\%).   

\begin{figure}[tbph]
\centering
\begin{subfigure}{.5\textwidth}
  \centering
  \includegraphics[width=.95\linewidth]{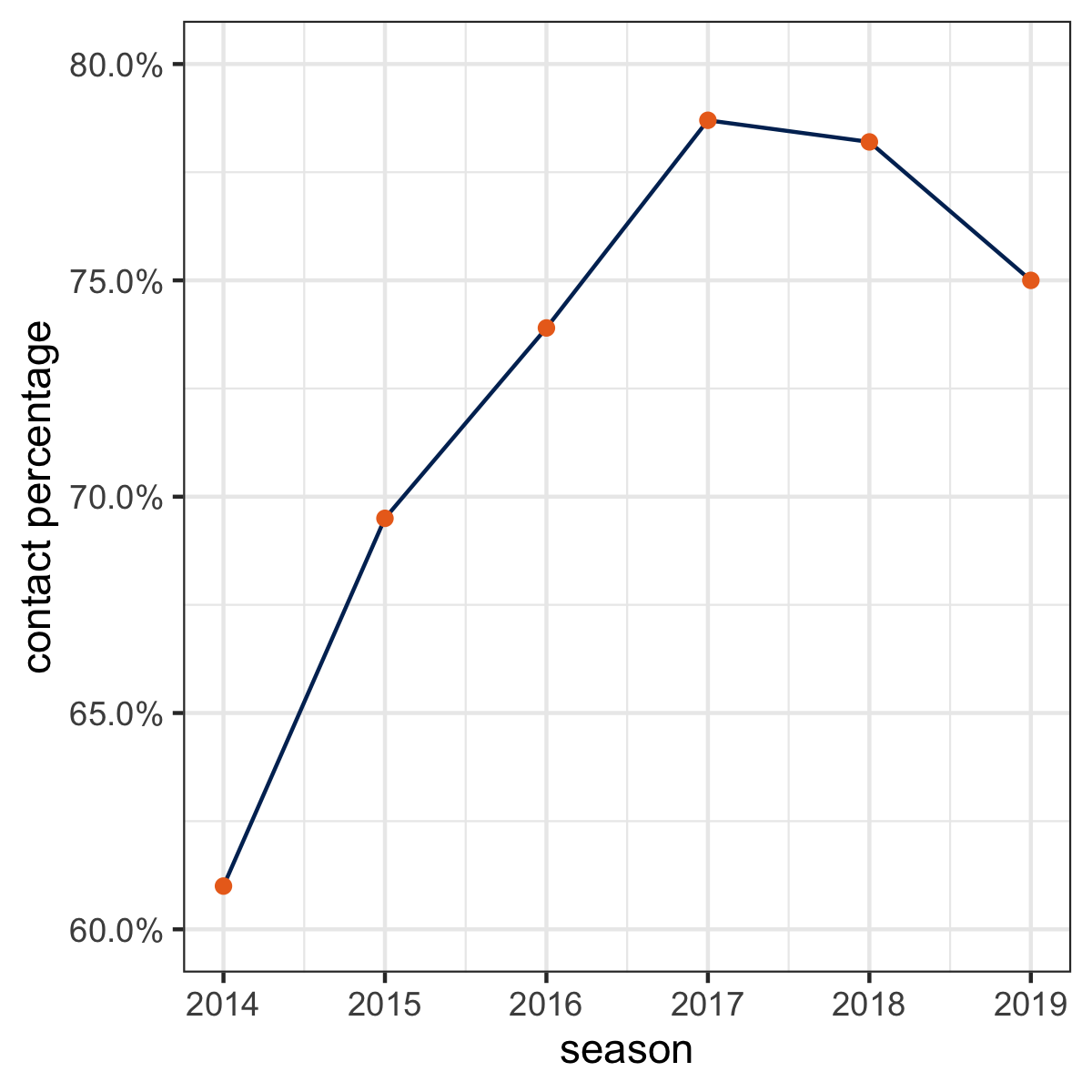}
  \caption{Contact Probability}
  \label{fig:contact_pct}
\end{subfigure}%
\begin{subfigure}{.5\textwidth}
  \centering
  \includegraphics[width=.95\linewidth]{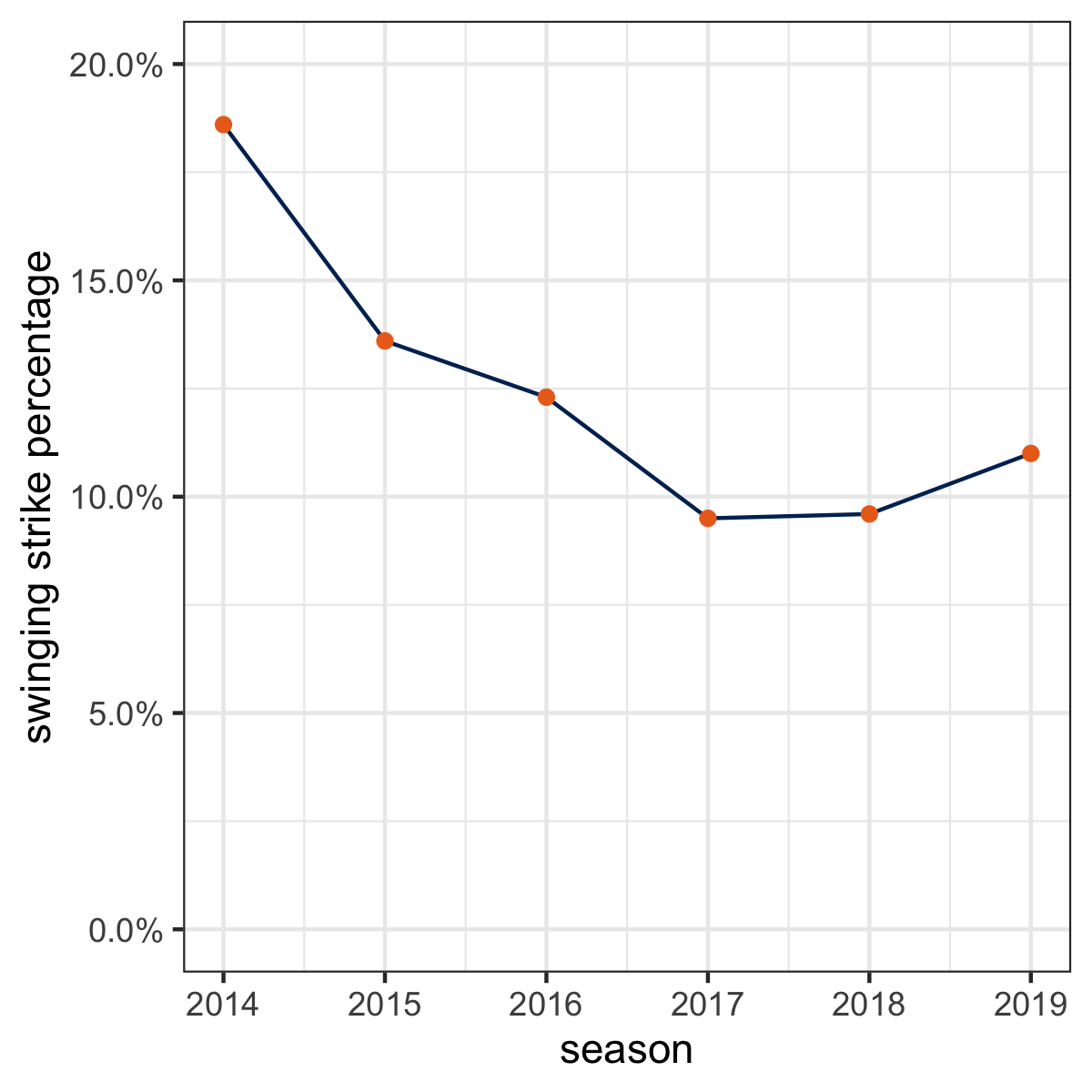}
  \caption{Swinging Strike Percentage}
  \label{fig:sw_strike}
\end{subfigure}
\caption{The percentage of swings in which George Springer made contact over each year in his career is given in panel (a). Swinging strike percentage over the same time period is shown in panel (b).}
\label{fig:springer_percentages}
\end{figure}

Using data obtained from \citet{Fangraphs}, Figures \ref{fig:contact_pct} and \ref{fig:sw_strike} show George Springer's contact percentage and swinging strike percentage, respectively, over the course of his career. Interestingly, Springer's contact percentage on swings over the course of his career, his two highest years of contact percentage were in 2017 and 2018 (78.7\% and 78.5\%, respectively), which are the two years that the Astros were accused of stealing signs. We see a similar pattern on swinging strike percentage in that Springer's career lows occur in 2017 and 2018 (9.5\% and 9.6\%), followed by an increase in 2019. Neither of these plots on their own are evidence of cheating as there could be numerous alternative explanations for this drop in swinging strike rate/increase in contact rate. However, these plots are consistent with evidence presented above that Springer was making contact at much higher rates when swinging at off-speed pitches preceded by a bang.  


\subsection{Exit Velocity Model}
To fit the exit velocity model, we restricted the observations used in this model to only include instances where contact (as defined above) was made and a launch speed, or exit velocity, was recorded. This leaves 2272 observations for our final analysis.  

\begin{table}[ht]
\centering
\caption{The fixed-effects estimates for our model on exit velocity. The effects involving bangs in shown in bold font.}
\renewcommand{\arraystretch}{1.25}
\begin{tabular}{rrrrr}
  \toprule
 Term & Estimate & Std. Error & Z Statistic & p-value \\ 
  \hline
  Intercept & 76.00 & 0.94 & 80.75 & 0.00  \\ 
  CSP & 8.36 & 0.91 & 9.18 & 0.00\\ 
  $I_{\{\textrm{Fastball}\}}$ & 2.20 & 0.70 & 3.15 & 0.00165 \\ 
  $I_{\{Bang\}}$ & {\bf 2.39} & 1.05 & 2.27 & 0.0230 \\ 
   \bottomrule
   \label{table_exit_velo_model}
\end{tabular}
\end{table}
 
Coefficient estimates for the exit velocity model are given in Table \ref{table_exit_velo_model}. The coefficient estimate for indicator of banging on a can in this model is 2.386 (95\% CI: 0.334, 4.451). Therefore, we estimate that when a batter makes contact with ball on a pitch preceded by bangs their exit velocity is 2.386 miles per hour greater on average than on pitches that were not preceded by a bang, when every other variable is held constant. To put this in perspective, a ball hit at 100 miles per hour at a launch angle of 30 degrees will travel roughly 385.3 feet before it hits the ground \citep{Nathan2020}. A ball hit with the same launch angle, but with an exit velocity of 102.38 (i.e., 2.386 miles per hour more) will travel 397.9 feet before hitting the ground, or 12.6 additional feet. This is the difference between a long fly ball to straight away center field (likely an out) and a home run at Fenway Park, where a home run is 389' 9'' inches to center field.




\section{Conclusion and Discussion}
\label{sec:conc}
In this manuscript, we examined the effects of sign stealing by the Houston Astros during the 2017 Major League Baseball season. We first verified that the presence of banging on a trashcan prior to a pitch was indeed related to the type of pitch being thrown. That is, do the bangs indicate that an off-speed pitch is likely to be thrown. The results are presented in Table \ref{table_pitch_group_vs_bang}. Next, we showed that the presence of the banging was significantly related to the probability of an Astros batter swinging, however, we did not control for potential confounding factors. 

In order to control for additional covariates, we used a series of generalized linear mixed effects models to control for known factors that potentially affect each of the outcomes considered here. Specifically, we modeled the probability that a player swings at a pitch, followed by modeling the probability of contact given a swing, and finally a model looking at exit velocity given contact. The three main findings of our paper are that the presence of bangs made it {\em less} likely that a player would swing at a pitch, {\em more} likely that a player would make contact with a off-speed pitch given that he swung, and {\em increased} the average exit velocity given that a player swung and made contact. Our findings are summarized below.

\begin{enumerate}
    \item The odds of swinging at a pitch were about 27.5\% lower when a bang was present prior to the pitch, OR 0.725 (95\% CI: (0.618, 0.850). 
    \item Given a swing, the odds of making contact with a breaking ball were about 80\% higher when a bang was present prior to the pitch, OR 1.805 (95\% CI: 1.342, 2.675). Note that the effect of bangs on fastball was not significant.
    \item Given a swing and contact, exit velocity was increased by an average of 2.386 (95\% CI:  0.334, 4.451) miles per hour when a bang was present.  
\end{enumerate}

In addition, we found that there was quite a bit of variability in how much the banging aided players in making contact with the ball given a swing. A particularly notable example is that George Springer was found to make contact on  swings of off-speed pitches at much higher rates when a bang was present relative to the same pitch with no bang, estimated OR 3.810 (95\% CI: 2.042, 12.864). 

In closing, we emphasize that these data and the results of our modeling efforts show that the effect of the Astros stealing the pitching sign significantly impacted their team's on-field performance. And while the effects were shown to differ from player to player, the overall impact on the game itself is undeniable -- the Astros were beneficiaries of their sign-stealing scheme and went on to win the 2017 World Series. Given the evidence presented here, we would argue that a cheater may indeed prosper, and occasionally even win a World Series. 

\section*{Acknowledgements}

The authors gratefully acknowledge Harry Pavlidis and all the members of the Baseball Prospectus Stats Slack channel, Tony Adams for collecting and disseminating the bangs data set, Tim P. Levine for the title suggestion, and Scott Sibbel for early comments and suggestions. 

\bigskip
\begin{center}
{\large\bf SUPPLEMENTARY MATERIAL}
\end{center}

All of the code and data related to this manuscript is publicly available on github at https://github.com/gjm112/Astros\_sign\_stealing

\bibliographystyle{agsm}
\bibliography{citations}

\end{document}